\documentstyle[sprocl]{article}
\input epsf.tex
\def\DESepsf(#1 width #2){\epsfxsize=#2 \epsfbox{#1}}
\def\be{\begin{equation}}
\def\ee{\end{equation}}
\def\bea{\begin{eqnarray}}
\def\eea{\end{eqnarray}}

\begin{document}
\title{
CP Violation in the SM and Beyond in Hadronic B Decays
}
\author{
Xiao-Gang He~\footnote{e-mail: hexg@phys.ntu.edu.tw}
}
\address{
Department of Physics, National Taiwan University,
Taipei, Taiwan 10764, R.O.C.
}
\author{  }
\address{ }
\maketitle
\begin{abstract}
Three different methods, using
$B_d\to J/\psi K_S$, $J/\psi K_S \pi^0$, $B_d\to K^-\pi^+, \pi^+\pi^-$
and $B_u\to K^- \pi^0, \bar K^0 \pi^-, \pi^-\pi^0$, to 
extract hadronic model independent 
information about new physics are discussed in this talk. 
\end{abstract}

\noindent
{\bf 1. Introduction}

In this talk I discuss three methods, using
$B_d\to$ $ J/\psi K_S$, $J/\psi K_S \pi^0$~\cite{5}, 
$B_d\to K^-\pi^+, \pi^+\pi^-$~\cite{6}
and $B_u\to K^- \pi^0, \bar K^0 \pi^-, \pi^-\pi^0$~\cite{7,8}, to 
extract hadronic model independent
information about the SM and models beyond.

The SM effective Hamiltonian respossible for hadronic decays is 
knownRef.~\cite{9}.
When going beyond the SM, there are new contributions. 
I will take three models beyond the SM for illustrations.
\\

\noindent
{\bf Model i): R-parity violation model}

In R-parity violating supersymmetric (SUSY) models, there are
new CP violating phases. Here I consider the effects
due to, $L = (\lambda''_{ijk}/2) U^{ci}_R D^{cj}_RD^{ck}_R$,
 R-parity violating 
interaction. 
Exchange of
$\tilde d_i$ squark 
can generate the following effective Hamiltonian at tree level,
$H_{eff} =(4G_F/ \sqrt{2})$ $V_{fb}^*V_{fq}$ $c^{f(q)}[O^{f(q)}_1(R) 
- O^{f(q)}_2(R)]$. Here
$O^{f(q)}_1(R) = \bar f\gamma_\mu R f \bar q \gamma^\mu R b$ and
$O^{f(q)}_2(R)$ $= \bar f_{\alpha}\gamma_\mu R f_{\beta} \bar q_{\beta} 
\gamma^\mu R b_{\alpha}$. The operators $O_{1,2}(R)$ have 
the opposite chirality
as those of the tree operators $O_{1,2}(L)$ in the SM.
The coefficients
$c^{f(q)}$ with QCD corrections are given by 
$(c_1-c_2)(\sqrt{2}/(4G_F V_{fb}V_{fq}^*))
(-\lambda''_{fqi}\lambda''^*
_{f3i}/2m^2_{\tilde d^i})$.
Here $m_{\tilde d^i}$ is the squark mass.
$B_u\to \pi^-\bar K^0$ and $B_u\to K^-\bar K^0$ data constrain 
$|c^{c(q)}|$ to be less than $O(1)$~\cite{10}.
The new contributions can be larger than the SM ones.
I will take the values to be 10\% 
of the corresponding values for the SM
with arbitrary phases $\delta^{f(q)}$ for later discussions.
\\

\noindent
{\bf Model ii): SUSY with large gluonic dipole interaction}

In SUSY models with R-parity conservation, potential large 
contributions to B decays may come from gluonic dipole interaction 
$c_{11}^{new}$ by exchanging gluino at loop
level with left- and right-handed squark mixing. 
$c_{11}^{new}$ is
constrained by experimental data from $b\to s\gamma$ which, however, still 
allows $c_{11}^{new}$ to be as large as
 three times of the SM contribution in magnitude
with an arbitrary CP violating phase $\delta_{dipole}$~\cite{11}.
I will take $c_{11}^{new}$ to be 3 times of the SM value with an arbitrary
$\delta_{dipole}$.
\\

\noindent
{\bf Model iii): Anomalous gauge boson couplings}

Anomalous gauge boson couplings can modify
the Wilson Coefficients of the SM ones with the same CP violating source 
as that for the SM~\cite{8,12}. 
The largest contribution may come from the $WWZ$ anomalous coupling
$\Delta g^Z_1$. 
LEP data constrain~\cite{13} $\Delta g_1^Z$ to be within 
$-0.113 <\Delta g_1^Z < 0.126$ at the 95\% c.l.. The resulting Wilson 
Coefficients can be
very different from those in the SM. 
\\

\noindent
{\bf 2. Test new physics and $\sin 2\beta$ from
$B\to J/\psi K_S, J/\psi K_S \pi^0$}
  
The usual $CP$ violation measure for $B$ decays to CP eigenstates
is ${\rm Im\,}\xi = {\rm Im\,}\{(q/p)(A^*\bar A/\vert A\vert ^2)\}$,
where $q/p = e^{-2i\phi_B}$ is from $B^0$--$\bar B^0$ mixing,
while $A$, $\bar A$ are
$B$,  $\bar B$ decay amplitudes.
For $B\rightarrow J/\psi K_S$, the final state is $P$-wave hence $CP$ odd.
Setting the weak phase in the decay amplitude to be $2\phi_0=Arg(A/\bar A)$, 
one has
, ${\rm Im\,}\xi(B\rightarrow J/\psi K_S) = -\sin(2\phi_B + 2 \phi_0)
\equiv - \sin2\beta_{J/\psi K_S}$.

For $B\rightarrow J/\psi K^*\rightarrow J/\psi K_S \pi^0$,
the final state has both $P$-wave ($CP$ odd) and
$S$- and $D$-wave ($CP$ even) components.
If $S$- and  $D$-wave have a common weak phase $\tilde \phi_1$ and 
P-wave has a weak phase $\phi_1$~\cite{5},
\begin{eqnarray}
{\rm Im\,}\xi(B\rightarrow J/\psi K_S \pi^0)
&=& {\rm Im\,} \{ e^{-2i\phi_B} [ e^{-2i\phi_1}\vert P \vert ^2
              -e^{-2i\tilde \phi_1}(1-\vert P \vert ^2)]\}\nonumber\\
&\equiv& (1-2\vert P \vert ^2)\sin2\beta_{J/\psi K_S\pi^0},
\end{eqnarray}
where $|P|^2$ is the fraction of P-wave component.
In the SM
one has $\phi_B = \beta$ and $2\phi_0=2\phi_1 (2\tilde \phi_1)
= Arg[\{V_{cb}V_{cs}^*
\{c_1 +a(a') c_2\}\}/\{V_{cb}^*V_{cs}\{c_1+a(a')c_2\}\}] = 0$, 
in the Wolfenstein phase convention. Here $a$ and $a'$ are 
parameters which indicate the relative contribution
from $O_{2}^{c(s)}(L)$ compared with $O_1^{c(s)}(L)$ for the P-wave and 
(S-, D-) wave. In the factorization approximation $1/a = 1/a' =N_c$
(the number of 
colors).
Therefore
$\sin2\beta_{J/\psi K_S}= \sin2\beta_{J/\psi K_S\pi^0}= \sin2\beta$.
$|P|^2$ has been measured with a small value~\cite{14}
$0.16\pm 0.08\pm 0.04$ by CLEO which
implies that the measurement of $\sin 2\beta$ using $B\to J/\psi K_S \pi^0$
is practical although
there is a dilution factor of 30\%.

When one goes beyond the SM,
$\sin2\beta_{J/\psi K_S}= \sin2\beta_{J/\psi K_S\pi^0}$ is not necessarily
true.
Let us now analyze the possible values for 
$\Delta \sin 2\beta \equiv
\sin2\beta_{J/\psi K_S}-\sin2\beta_{J/\psi K_S\pi^0}$. 
Because $B\to J/\psi K_S,
J/\psi K^*$ are tree dominated processes, Models ii) and iii) would not
change the SM predictions significantly. $\Delta \sin 2\beta$ is not
sensitive to new physics in Models ii) and iii). However, for Model i),
the contributions can be large. 
The weak phases are given by 
$2\phi_0 = 2\phi_1(2\tilde \phi_1) = Arg[
\{V_{cb}V_{cs}^*\{c_1+a c_2 +(-)
 c^{c(s)}\{1-a(a')\}\}\}
/\{V_{cb}^*V_{cs}\{c_1+a c_2 +(-)
 c^{c(s)*}\{1-a(a')\}\}\}]$.
Taking the new contributions to be 10\% of the
SM ones, one obtains $\phi_0=\phi_1 \approx -\tilde \phi_1 \approx
0.1\sin\delta^{c(s)}$. From this,
$\Delta \sin 2\beta \approx 4((1-|P|^2)/(1-2|P|^2))\cos 2\phi_B 
(0.1\sin\delta^{c(s)})\approx 0.5 \cos 2\phi_B \sin\delta^{c(s)}$. 
$\phi_B$ may be different
from the SM one due to new contributions. Using the central
value $\sin 2\beta_{J/\psi K_S} = 0.91$ measured from CDF and ALEPH~\cite{1}, 
$\Delta \sin 2\beta \approx 0.2\sin \delta^{c(s)}$ which 
can be as large as 0.2. Such a large difference
can be measured at B factories. Information about new CP violating
phase $\delta^{c(s)}$ can be obtained.
\\

\noindent  
{\bf 3. Test new physics and rate differences between $B_d\to \pi^+ K^-, 
\pi^+\pi^-$}

I now show that hadronic model independent
information about CP violation can be obtained using SU(3) analysis for
rare hadronic B decays.

The SM operators $O_{1,2}$, $O_{3-6, 11,12}$, and $O_{7-10}$ for 
rare hadronic B decays transform under SU(3)
symmetry as $\bar 3_a + \bar 3_b +6 + \overline {15}$,
$\bar 3$, and $\bar 3_a + \bar 3_b +6 + \overline {15}$, respectively. 
These properties enable one to 
write the decay amplitudes for $B\to PP$ in only a few SU(3) invariant 
amplitudes.
When small annihilation contributions are neglected, one has

\begin{eqnarray}
&&A(B_d\to \pi^+\pi^-) = V_{ub}V_{ud}^* T + V_{tb}V_{td}^* P,\nonumber\\
&&A(B_d\to \pi^+ K^-) = V_{ub}V_{us}^* T + V_{tb}V_{ts}^* P.
\nonumber
\end{eqnarray}
From above one obtains, $\Delta(\pi^+\pi^-) =- \Delta(\pi^+ K^-)$.
This non-trivial equality dose not
depend on detailed models for hadronic physics 
and provides test for the SM~\cite{6}. 
Including SU(3) breaking effect from factorization calculation, one
has, $\Delta(\pi^+\pi^-) \approx - {f_\pi^2\over f_K^2}
\Delta(\pi^+ K^-)$. Although there is correction, the relative
sign is not changed.

When going beyond the SM, there are new CP violating phases leading to
violation of the equality above. 
For example Models i) and ii) can alter the equality significantly, while
Model iii) can not because the CP violating source is
the same as that in the SM. To illustrate how the situation is changed in Models
i) and ii), I calculate the normalized asymmetry 
$A_{norm}(PP)=$$\Delta(PP)/\Gamma(\pi^+K^-)$ using factorization 
approximation following Ref.~\cite{hhy}.
The new effects may come in such a way that only $B_d\to \pi^+ K^-$ is 
changed but not
$B_d \to \pi^+\pi^-$. This scenario leads to maximal violation of the equality 
discussed here. 

The results are shown in Figure 1. 
The solid curve is the SM prediction for $A_{norm}(\pi^+ K^-)$ 
as a function of $\gamma$. For $\gamma_{best}$ $A_{norm}(\pi^+ K^0) 
\approx 10\%$. 
It is clear from Figure 1 that
within the allowed range of the parameters, new physics effects can 
dramatically violate the equality discussed above. 
\\

\begin{figure}[htb]
\centerline{ \DESepsf(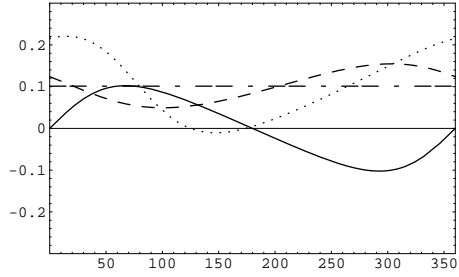 width 6 cm) }
\caption {$A_{norm}(\pi^+ K^-)$ vs. 
phases $\gamma$, $\delta^{u(s)}$ and $\delta_{dipole}$ for 
the SM (solid), 
Models i) (dashed) and ii) (dotted). 
For Models i) and ii), $\gamma = 59.9^\circ$ is used.
$A_{norm}(\pi^+ K^-) =$ $ -(f_\pi/f_k)^2
A_{norm}(\pi^+\pi^-)$ is satisfied in the SM. For Models i) and ii)
$-(f_\pi/f_k)^2A_{norm}(\pi^+\pi^-)$ is approximately the same as that in the 
SM (dot-dashed curve).}
\end{figure}

\noindent
{\bf 4. Test new physics and SU(3) relation for $B_u\to \pi^-\bar K^0,
\pi^0 K^-, \pi^0\pi^-$}

I now discuss another method which provides important information about
new physics using $B_u\to \pi^- \bar K^0, \pi^0 K^-, \pi^0\pi^-$.
Using SU(3) relation and factorization estimate  for
the breaking effects, one 
obtains~\cite{7,8}

\begin{eqnarray}
A(B_u \to \pi^- \bar K^0) + \sqrt{2}A(B_u\to \pi^0 K^-)
&=&\epsilon A(B_u \to \pi^-\bar K^0)e^{i\Delta \phi}
(e^{-i\gamma}-\delta_{EW}),\nonumber\\
\delta_{EW} = -{3\over 2} {|V_{cb}||V_{cs}|\over |V_{ub}||V_{us}|}{c_{9}+c_{10}\over c_1+c_2},
&&\epsilon = \sqrt{2} {|V_{us}|\over |V_{ud}|} {f_K\over f_\pi}
{|A(\pi^+\pi^0)|\over |A(\pi^+ K^0)|},
\nonumber
\end{eqnarray}
where $\Delta \phi$ is the difference of the final state rescattering phases
for $I=3/2,1/2$ amplitudes. 
For $f_K/f_\pi = 1.22$ and 
$Br(B^\pm \to \pi^\pm \pi^0) = (0.54^{+0.21}_{-0.20}\pm 0.15)\times 10^{-5}$~\cite{4a}, 
one obtains $\epsilon = 0.21 \pm 0.06$.

Neglecting small tree contribution to $B_u\to \pi^- \bar K^0$, one obtains
\begin{eqnarray}
\cos\gamma =\delta_{EW} -
{(r^2_++r^2_-)/2 -1 - \epsilon^2(1-\delta_{EW}^2) 
\over 2 \epsilon (\cos \Delta \phi
+ \epsilon \delta_{EW})},
\;\;r^2_+-r^2_- = 4\epsilon \sin \Delta \phi \sin \gamma,
\nonumber
\end{eqnarray}
where
$r_\pm^2 = 4Br(\pi^0 K^\pm)/[Br(\pi^+ K^0)
+ Br(\pi^- \bar K^0)] = 1.33\pm 0.45$~\cite{4a}. 

It is interesting to note that although the above equation is complicated, 
bound on $\cos\gamma$ can be 
obtained~\cite{8}. 
For $\Delta = (r^2_+ + r^2_-)/2 -1-\epsilon^2(1-\delta_{EW}^2)\ge (\le)>0$, 
we have

\begin{eqnarray}
\cos\gamma \le (\ge) \delta_{EW}- {\Delta\over 2\epsilon (1+\epsilon \delta_{EW})}, \;\;
\mbox{or}\;\;\;
\cos\gamma \ge (\le)\delta_{EW}-{\Delta \over 2\epsilon (-1+\epsilon \delta_{EW})}.
\label{bound}
\end{eqnarray}

The bounds on $\cos\gamma$ as a function of $\delta_{EW}$ are shown in Fig. 2 
by the solid curves
for three representative cases: 
a) Central values for $\epsilon$ and $r^2_\pm$;
b) Central values for $\epsilon$ and $1 \sigma$ upper bound $r^2_\pm=1.78$;
and c) Central value for $\epsilon$ and $1 \sigma$ lower 
bound $r^2_\pm = 0.88$.
The bounds with $|\cos\gamma| \le 1$ for a), b) and c) are indicated by   
the curves (a1, a2), (b) 
and (c1, c2), respectively.
For cases  a) and c) there are two allowed 
regions, 
the regions below (a1, c1) and the regions above (a2, c2).
For case b) the allowed range is below (b).

\begin{figure}[htb]
\centerline{ \DESepsf(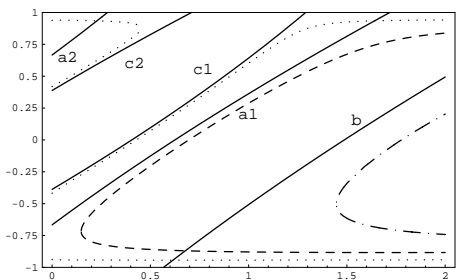 width 6 cm) }
\caption {$\cos\gamma$ vs. $\delta_{EW}$. 
The solutions for the cases a), b) and c) are
indicated by the dashed, dot-dashed and dotted curves 
respectively.}
\end{figure}

One also has
\begin{eqnarray}
(1-\cos^2\gamma) [ 1-({\Delta \over 2\epsilon(\delta_{EW} -\cos \gamma)}
-\epsilon \delta_{EW})^2] - {(r^2_+-r^2_-)^2\over 16 \epsilon^2} = 0.
\end{eqnarray}
To have some idea about the details, I analyze 
the solutions of $\cos \gamma$ as a function of $\delta_{EW}$
for the three cases discussed earlier 
with a given value for the asymmetry $A_{asy} = (r^2_+-r^2_-)/(r^2_++r^2_-)
=15\%$.

In the SM for $r_v=|V_{ub}|/|V_{cb}| = 0.08$ and $|V_{us}| = 0.2196$, 
$\delta_{EW} = 0.81$. The central values for $r_\pm$ and $\epsilon$ prefers
$\cos\gamma < 0$ which is different from the result obtained in Ref.~\cite{2} 
by fitting other
data. The parameter $\delta_{EW}$ is sensitive to new physics in the 
tree and electroweak sectors. Model i) has large corrections to the tree level
contributions. However, the 
contribution is proportional to the sum of the 
coefficients of operators $O_{1,2}^{u(s)}(R)$ which is zero in Model i). 
The above method does not provide information about new physics due to
Model i). This method would not provide information about new physics due to
Model ii) neither because the gluonic dipole interaction transforms 
as $\bar 3$ which
does not affect $\delta_{EW}$. Model iii) can have large effect on 
$\delta_{EW}$. In this model $\delta_{EW} = 0.81(1+ 4.33\Delta g_1^Z)$ which 
can vary in the range $0.40\sim 1.25$. 

For case a), in the SM $\cos\gamma <0.18$ which is inconsistent 
with $\cos\gamma_{best}\approx 0.5$ from other fit~\cite{2}.
In Model iii) $\cos \gamma$ can be consistent with $\cos\gamma_{best}$.
For case b), $\cos\gamma$ is less than zero in both the 
SM and Model iii).
If this is indeed the case, other types of new physics is needed.
For case c) $\cos \gamma$ can be close to $\cos\gamma_{best}$ 
for both the SM and Model iii).
\\

\noindent
{\bf 5. Conclusion}

From discussions in previous sections, it is clear that using
$B_d\to J/\psi K_S$, $J/\psi K_S \pi^0$, $B_u\to \pi^- K^+, \pi^+\pi^-$
and $B^-\to \pi^0 K^-, \pi^- \bar K^0, \pi^0\pi^-$
important information free from uncertainties in hadronic physics about 
the Standard Model and models beyond 
can be obtained. These analyses should be carried out at B factories.
\\

\noindent
{\bf Acknowledgements} 

This work was partially supported by National Science Council of
R.O.C. under grant number  NSC 89-2112-M-002-016. I thank
Deshpande, Hou, Hsueh and Shi for collaborations on materials presented
in this talk.

\end{document}